\documentclass[conference]{IEEEtran}
\IEEEoverridecommandlockouts
\usepackage{cite}
\usepackage{amsmath,amssymb,amsfonts}
\usepackage{algorithmic}
\usepackage{graphicx}
\usepackage{textcomp}
\usepackage{xcolor}
\def\BibTeX{{\rm B\kern-.05em{\sc i\kern-.025em b}\kern-.08em
    T\kern-.1667em\lower.7ex\hbox{E}\kern-.125emX}}

\usepackage[english]{babel}
\usepackage[nolist]{acronym}

\begin{acronym}

\acro{3DES}{Triple-DES}

\acro{AES}{Advanced Encryption Standard}
\acro{ACT-R}{Adaptive Control of Thought-Rational}
\acro{ALU}{Arithmetic Logic Unit}
\acro{API}{Application Programming Interface}
\acro{ARX}{Addition Rotation XOR}
\acro{ASIC}{Application Specific Integrated Circuit}
\acro{ASIP}{Application Specific Instruction-Set Processor}
\acro{AS}{Active Serial}

\acro{BDD}{Binary Decision Diagram}
\acro{BGL}{Boost Graph Library}
\acro{BNF}{Backus-Naur Form}
\acro{BRAM}{Block-Ram}

\acro{CBC}{Cipher Block Chaining}
\acro{CFB}{Cipher Feedback Mode}
\acro{CFG}{Control Flow Graph}
\acro{CLB}{Configurable Logic Block}
\acro{CLI}{Command Line Interface}
\acro{COFF}{Common Object File Format}
\acro{CPA}{Correlation Power Analysis}
\acro{CPU}{Central Processing Unit}
\acro{CRC}{Cyclic Redundancy Check}
\acro{CTR}{Counter}

\acro{DC}{Direct Current}
\acro{DES}{Data Encryption Standard}
\acro{DFA}{Differential Frequency Analysis}
\acro{DFT}{Discrete Fourier Transform}
\acro{DLL}{Dynamic Link Library}
\acro{DMA}{Direct Memory Access}
\acro{DNF}{Disjunctive Normal Form}
\acro{DPA}{Differential Power Analysis}
\acro{DSO}{Digital Storage Oscilloscope}
\acro{DSP}{Digital Signal Processing}
\acro{DUT}{Design Under Test}

\acro{ECB}{Electronic Code Book}
\acro{ECC}{Elliptic Curve Cryptography}
\acro{EEPROM}{Electrically Erasable Programmable Read-only Memory}
\acro{EMA}{Electromagnetic Emanation}
\acro{EM}{electro-magnetic}

\acro{FFT}{Fast Fourier Transformation}
\acro{FF}{Flip-Flop}
\acro{FI}{Fault Injection}
\acro{FIR}{Finite Impulse Response}
\acro{FPGA}{Field Programmable Gate Array}
\acro{FSIQ}{Full Scale IQ}
\acro{FSM}{Finite State Machine}

\acro{GUI}{Graphical User Interface}
\acro{GDPR}{General Data Protection Regulation}

\acro{HDL}{Hardware Description Language}
\acro{HD}{Hamming Distance}
\acro{HF}{High Frequency}
\acro{HRE}{Hardware Reverse Engineering}
\acro{HSM}{Hardware Security Module}
\acro{HW}{Hamming Weight}

\acro{IC}{Integrated Circuit}
\acro{IO}[I/O]{Input/Output}
\acro{IOB}{Input Output Block}
\acro{IOT}[IoT]{Internet of Things}
\acro{IP}{Intellectual Property}
\acro{ISA}{Instruction Set Architecture}
\acro{IV}{Initialization Vector}

\acro{JTAG}{Joint Test Action Group}

\acro{KAT}{Known Answer Test}

\acro{LFSR}{Linear Feedback Shift Register}
\acro{LSB}{Least Significant Bit}
\acro{LUT}{Look-Up Table}

\acro{MAC}{Message Authentication Code}
\acro{MIPS}{Microprocessor without Interlocked Pipeline Stages}
\acro{MMIO}{Memory Mapped \acl{IO}}
\acro{MSB}{Most Significant Bit}
\acro{MUX}{Multiplexer}

\acro{NASA}{National Aeronautics and Space Administration}
\acro{NSA}{National Security Agency}
\acro{NVM}{Non-Volatile Memory}

\acro{OFB}{Output Feedback Mode}
\acro{OISC}{One Instruction Set Computer}
\acro{ORAM}{Oblivious Random Access Memory}
\acro{OS}{Operating System}

\acro{PAR}{Place-and-Route}
\acro{PCB}{Printed Circuit Board}
\acro{PC}{Personal Computer}
\acro{PR}{Perceptual Reasoning}
\acro{PS}{Processing Speed}
\acro{PUF}{Physical Unclonable Function}

\acro{QCM}{Questionnaire on Current Motivation}

\acro{RISC}{Reduced Instruction Set Computer}
\acro{RNG}{Random Number Generator}
\acro{ROM}{Read-Only Memory}
\acro{ROP}{Return-oriented Programming}
\acro{RTL}{Register Transfer Level}

\acro{SCA}{Side-Channel Analysis}
\acro{SHA}{Secure Hash Algorithm}
\acro{SNR}{Signal-to-Noise Ratio}
\acro{SPA}{Simple Power Analysis}
\acro{SPI}{Serial Peripheral Interface Bus}
\acro{SRAM}{Static Random Access Memory}

\acro{UART}{Universal Asynchronous Receiver Transmitter}
\acro{UHF}{Ultra-High Frequency}

\acro{VC}{Verbal Comprehension}
\acro{VLSI}{Very-Large-Scale Integration}
\acro{VHDL}{Very High Speed Integrated Circuit Hardware Description Language}

\acro{WAIS-IV}{Wechsler Adult Intelligence Scale}
\acro{WISC}{Writeable Instruction Set Computer}
\acro{WM}{Working Memory}

\acro{XDL}{Xilinx Description Language}
\acro{XTS}{XEX-based Tweaked-codebook with ciphertext Stealing}

\end{acronym}

\usepackage[utf8]{inputenc}
\usepackage[babel=true]{csquotes}
\usepackage{array}
\usepackage{comment}
\usepackage{hyperref}
\usepackage{listings} 
\usepackage{makecell}
\usepackage{microtype}
\usepackage{tikz}
\usetikzlibrary{arrows}
\usetikzlibrary{automata}
\usetikzlibrary{circuits.logic.US}
\usetikzlibrary{decorations.text}
\usetikzlibrary{matrix}
\usetikzlibrary{shadows}
\usepackage{url}
\usepackage{xspace}

\ifCLASSOPTIONcompsoc
\usepackage[caption=false,font=normalsize,labelfont=sf,textfont=sf]{subfig}
\else
\usepackage[caption=false,font=footnotesize]{subfig}
\fi

\newcommand{\HAL}{\textnormal{\textsf{HAL}}\xspace}

\definecolor{monokai_background}{RGB}{39, 40, 34}
\definecolor{monokai_blue}{RGB}{137, 189, 255}
\definecolor{monokai_green}{RGB}{166, 226, 42}
\definecolor{monokai_red}{RGB}{249, 38, 114}

\addto\extrasenglish{

}

\addto\captionsenglish{\renewcommand{\figurename}{Fig.}}
\begin{document}
	
\title{Promoting the Acquisition of Hardware Reverse Engineering Skills}

\author{\normalsize{Carina~Wiesen$^1$$^,$$^2$, Steffen~Becker$^2$, Nils~Albartus$^2$, Christof~Paar$^2$, and~Nikol~Rummel$^1$$^,$$^2$}\\
	\normalsize{$^1$\textit{Institute of Educational Research} and $^2$\textit{Horst G\"ortz Institute for IT Security}}\\
	\normalsize{\textit{Ruhr University Bochum}}\\
	\normalsize{Bochum, Germany}\\
	\normalsize{\{carina.wiesen, steffen.becker, nils.albartus, christof.paar, nikol.rummel\}@rub.de}\\\\

	\thanks{This work was supported in part through DFG Excellence Strategy grant 39078197 (EXC 2092, CASA), ERC grant 695022 and NSF grant CNS-1563829.}
}

\maketitle


\begin{abstract}
This full research paper focuses on skill acquisition in \ac{HRE} -- an important field of cyber security.
\ac{HRE} is a prevalent technique routinely employed by security engineers (i) to detect malicious hardware manipulations, (ii) to conduct \acs{VLSI} failure analysis, (iii) to identify \acs{IP} infringements, and (iv) to perform competitive analyses.
Even though the scientific community and industry have a high demand for \ac{HRE} experts, there is a lack of educational courses.
We developed a university-level \ac{HRE} course based on general cognitive psychological research on skill acquisition, as research on the acquisition of \ac{HRE} skills is lacking thus far. 
To investigate how novices acquire \ac{HRE} skills in our course, we conducted two studies with students on different levels of prior knowledge. Our results show that cognitive factors (e.g., working memory), and prior experiences (e.g., in symmetric cryptography) influence the acquisition of \ac{HRE} skills. We conclude by discussing implications for future \ac{HRE} courses and by outlining ideas for future research that would lead to a more comprehensive understanding of skill acquisition in this important field of cyber security.
\end{abstract}



\section{Introduction}
\label{fie:section:introduction}

In an increasingly digital world, individuals, industry, and governments wrestle with major cyber-security challenges posed by numerous and increasingly frequent cyber attacks at the hardware, software, and network level. Hardware components serve as the basis of trust in virtually any computing system by ensuring its security, integrity, and reliability.
Due to the globalized fabrication processes through which they are produced, however, \acp{IC} are vulnerable to attacks such as malicious manipulations or the insertion of hardware Trojans~\cite{ieee:2014:rostami,ches:2013:becker}.
The deployment of manipulated hardware chips in critical infrastructure such as cellular networks and power grids or in sensitive applications (e.g., aerospace or military) is a major concern for a wide array of stakeholders~\cite{IEEE:2014:Guin, JET:2014:Guin}.

A common method to detect such malicious manipulations in \acp{IC} is called \acf{HRE}.
\textit{Reverse engineering} can be described as the process of retrieving information from anything man-made to understand its inner structures and workings~\cite{smc:1985:rekoff}.
\ac{HRE} is a multi-layered process which facilitates the security inspection of an (unknown) hardware design~\cite{ivsw:2017:fyrbiak}.
Specifically, \ac{HRE} is employed for the purpose of \ac{VLSI} failure analysis, detecting counterfeits, identifying IP violations, and searching for potentially implanted backdoors and hardware Trojans~\cite{ches:2009:torrance}.
At the same time, malign actors can utilize \ac{HRE} for illegitimate purposes such as IP fraud or the insertion of backdoors or hardware Trojans. 

The continuous evolution of a digital society shaped by a rapidly expanding \ac{IOT} and the proliferation of cyber-physical systems have created a high demand for security experts with a solid background in \ac{HRE}. Nevertheless, there is an almost complete lack of educational courses in the \ac{HRE} field and \ac{HRE} training happens almost entirely on the job~\cite{TALE:2018:Wiesen}. We developed an \ac{HRE} course based on cognitive psychological research on skill acquisition, as research on the acquisition of \ac{HRE} skills is lacking thus far. The course was first offered at a German university during the 17/18 winter term. The second iteration of the course was introduced at one German and one North American university during the 18/19 winter term. In order to evaluate if our course actually enables \ac{HRE} skill acquisition, we conducted two studies with students on different levels of prior knowledge in relevant topics. We argue that research on \ac{HRE} skill acquisition and how to best foster it through educational courses is essential for enhancing the development of future university-level programs and for addressing the high and unmet demand of \ac{HRE} experts. 
  
We aim to observe if students of our course are able to acquire \ac{HRE} skills by taking a first step towards closing the research gap of understanding how \ac{HRE} skills are acquired. Our research will help to define future goals for teaching and learning in this specific field of cyber security. \\
In summary, our contributions are to:
\begin{itemize}
	\item Illustrate the current lack of research on how skills in this important field of cyber security are acquired.
	\item Provide an overview of prior research from cognitive psychology on the acquisition of skills.
	\item Develop a course based on the findings from cognitive psychology to enhance skill acquisition in \ac{HRE}.
	\item Observe and evaluate if our course enables students to acquire \ac{HRE} skills. Therefore, we present our two studies by formulating research questions, expound upon our study design and methods, and present the results of our examination. 
	\item Discuss the findings of our research on \ac{HRE} skill acquisition by evaluating our course design and by providing recommendations for future courses and research studies. 
\end{itemize}


\section{Background}
\label{fie:section:background}

\subsection{Hardware Reverse Engineering}
\label{FIE:background:hre}
The term \textit{reverse engineering} refers to the processes of extracting knowledge or design information from anything man-made in order to comprehend its inner structure~\cite{smc:1985:rekoff}.
In the context of hardware security~\cite{book:hardware_obfuscation:chapter1}, security engineers (as well as malicious actors) are forced to employ different techniques to extract a gate-level netlist from a given \ac{IC} or \ac{FPGA}.
The analysis of the gate-level netlist marks the crucial step of \ac{HRE} enabling human reverse engineers to make sense of an unknown hardware design (e.g., to identify security vulnerabilities or security-circuitry for Trojan insertion~\cite{ivsw:2017:wallat})~\cite{ivsw:2017:fyrbiak}.
In the following we define the term gate-level netlist reverse engineering as an important cyber security skill. Additionally, we outline the current lack of research on \ac{HRE} skill acquisition.

\subsubsection{Gate-level Netlist Reverse Engineering}
During the hardware design process, synthesis tools convert \ac{RTL} descriptions of hardware designs into representations of the (Boolean) logic gates of the target gate library and their connectivity~\cite{Weste:2010:CVD:1841628}. Such representations are called gate-level netlists. 

During the different synthesis steps, valuable high-level information such as (1)~meaningful descriptive information (e.g., names and comments), (2)~boundaries of implemented modules, and (3)~module hierarchies is lost. In practice, this loss of information highly complicates the reverse engineering process~\cite{tect:2013:subramanyan}.

In real-world settings, analysts can obtain gate-level netlists in several scenarios: (1)~through chip-level reverse engineering  in the case of a given \ac{IC} (involving steps such as (i)~decapsulation, (ii)~delayering, (iii)~image acquisition, and (iv)~image processing)~\cite{ches:2009:torrance}; (2)~through bitstream reverse engineering in the case of \acp{FPGA} (involving steps such as (i)~bitstream extraction or interception, (ii)~bitstream file format reversing, and (iii)~bitstream conversion)~\cite{fpga:2008:note}; or (3)~by direct access at a foundry or through bribery or theft.

Human analysts conducting gate-level netlist reverse engineering are often supported by semi-automatable tools that enable algorithmic graph-based and machine learning methods as well as allow for detailed visual inspection for the purpose of structural and functional analyses~\cite{tect:2013:subramanyan,aspdac:2019:Wiesen}.
Since no fully automated tools for \ac{HRE} exist, human engineers are always involved in the process of gate-level netlist reversing.
\subsubsection{Lack of Research on \ac{HRE} Skill Acquisition}
\ac{HRE} specialists need to draw upon knowledge from various domains including
chip design and manufacturing,
image processing and machine learning techniques,
Boolean algebra, graph theory,
custom tooling, programming languages, \acp{HDL},
computer architectures, 
cryptography, and cryptanalysis.
There are currently very few trained \ac{HRE} specialists, and due to the lack of educational courses in the field, most have acquired their skills through on-the-job-training or even through their free time pursuits as hobbyists.
This ad-hoc training situation is unsatisfactory since there is growing demand in industry and government agencies for \ac{HRE} specialists, which should be met by structured university-level courses. Therefore, we developed an \ac{HRE} course which is based on cognitive psychology, since research on the acquisition of \ac{HRE} skills is missing. In the following, we first address relevant aspects of psychological research on skill acquisition. Second, we apply the psychological research findings in order to develop and present the \ac{HRE} course design.   

\subsection{Psychological Background}
\label{FIE:background:psychbackgr}
Even though industry and the scientific community have a high demand for engineers with \ac{HRE} expertise, there is a lack of systematic research on the acquisition of \ac{HRE} skills. We are aware of one relevant prior work on human factors in reverse engineering ~\cite{lee2013theory}. The authors focused on exploring human problem-solving processes in reverse engineering of simple Boolean systems in an artificial laboratory setting. Thus, this prior work does not compare to \ac{HRE} practice. Additionally, the authors did not include research on skill acquisition processes. In the following sections, we present relevant psychological literature on skill acquisition which build the foundation of our \ac{HRE} course design.

\subsubsection{Skill Acquisition}
\label{FIE:background:knowacqu}
According to the \ac{ACT-R}~\cite{anderson1982acquisition}, knowledge is represented in two ways~\cite{tenison2016modeling}: \textit{Declarative} knowledge which consists of facts (e.g., the control path is operationalized as a \ac{FSM}), and \textit{procedural} knowledge which consists of mappings of stages to actions (e.g., if the goal is to find an \ac{FSM}, then search for strongly connected components in the netlist). \ac{ACT-R} includes assumptions about transferring declarative knowledge (\textit{knowing that}) into procedural knowledge (\textit{knowing how}): knowledge is first acquired declaratively, and afterwards transformed into a procedural form~\cite{gobet2005chunking}.

The acquisition of declarative (verbal) and procedural (non-verbal) knowledge is supported by various learning processes~\cite{koedinger2012knowledge}. These in turn can be reinforced through specific forms of instruction and course structure~\cite{koedinger2012knowledge}, which we incorporated into the design of our \ac{HRE} tasks (see Section~\ref{fie:section:materials}). In the following, we briefly describe the learning processes for declarative and procedural knowledge.

First, \textit{understanding and sense-making processes} involve verbally-mediated and explicit processes in which students attempt to understand and reason. Understanding and sense-making processes are more deliberate, since students need to actively engage in understanding and reasoning~\cite{koedinger2012knowledge}. Second, \textit{memory and fluency-building processes} are defined as non-verbal learning processes which involve strengthening memory and compiling knowledge~\cite{koedinger2012knowledge}. Fluency building enables solving tasks and problems more efficiently, since knowledge is more strongly composed and automatically accessible~\cite{singer2005long}~\cite{skinner1996increasing}. Prior research has shown that people speed up with practice~\cite{tenison2016modeling}. In this context, the psychological construct behind ``speeding up'' is called fluency which is defined as the ability to quickly and accurately solve a problem~\cite{kilpatrick2001strands}. Prior findings have shown that students with high scores in fluency maintain their skills over time~\cite{singer2005long} and perform better on more complex tasks than students with lower fluency~\cite{skinner1996increasing}. Third, \textit{induction and refinement processes} are non-verbal processes and improve the accuracy of knowledge through generalization, categorization, discrimination, or causal induction~\cite{koedinger2012knowledge}. 

Additionally, individual differences and abilities (e.g., intelligence) play an important role in the development of broad and complex skills~\cite{kyllonen1989role}. Furthermore, motivation is often described as a central driver of devoting years to deliberate practice and learning~\cite{litzinger2011engineering}. A high level of motivation leads to more cognitive engagement, more learning, and higher levels of achievement~\cite{pintrich2003motivational}, and is therefore relevant to the development of a course on \ac{HRE} which supports students’ learning processes. 

In the following section, we use concepts described in existing literature on skill and knowledge acquisition and instructional principles which support these learning processes~\cite{koedinger2012knowledge} to describe the design of our course in \ac{HRE}.

\subsubsection{Psychological Research as the Foundation for \ac{HRE} Course Design}
\label{FIE:background:reforcd}
With the goal of designing the course to enhance the acquisition of \ac{HRE} skills, we acknowledged the distinction between declarative and procedural knowledge acquisition by assuming that knowledge is first acquired declaratively, and is then transformed into a procedural form~\cite{anderson1982acquisition}~\cite{gobet2005chunking}. Practically, we divided the course into a lecture phase (acquisition of declarative knowledge) and a practical phase (transformation into procedural knowledge). We developed the learning materials for and instructional principles behind the two phases based on verbally-mediated and non-verbal learning processes as proposed in~\cite{koedinger2012knowledge}. Additionally, we considered potential influences upon student motivation as described in the following section.

\subsubsection{Lecture Phase}
During the first six weeks of the course, students acquire declarative knowledge. This phase focuses on the acquisition of verbally-mediated facts, theories, and concepts related to the relevant fields of electrical engineering, Boolean algebra, and graph theory through two 90-minute lectures and one homework assignment per week. We apply the instructional principles of Prompted Self-Explanation~\cite{koedinger2012knowledge} and Accountable Talks~\cite{michaels2008deliberative}~\cite{koedinger2012knowledge} to support the learning processes of \textit{understanding and sense making} (Section~\ref{FIE:background:knowacqu}). We achieve robust learning of declarative knowledge through the integration of verbally-mediated exercises which encourage students to explain the steps of Worked Examples of \ac{HRE} to themselves and to share their solutions with other students in accountable discussions in tutorial sessions.
\subsubsection{Practical Phase}
Following the lecture phase, students participate in an eight-week practical study consisting of four \ac{HRE} problems (detailed descriptions in Section~\ref{fie:section:materials}). Our decision to organize the course in two phases reflects our assumption that the declarative knowledge imparted during the lecture phase is transformed into practical knowledge through the non-verbal learning processes of the practical phase. To support non-verbal \textit{memory and fluency-building processes} (Section~\ref{FIE:background:knowacqu}), we leverage the instructional principle of Spacing and Testing~\cite{pashler2007organizing}~\cite{koedinger2012knowledge} by directing students to practice recalling target task material over longer time intervals (two weeks per project) to enhance their long-term retention and improve their fluency in solving \ac{HRE} problems. 

Additionally, we include non-verbal processes that are associated with the \textit{induction and refinement processes} (Section~\ref{FIE:background:knowacqu}), through the incorporation of Worked Examples~\cite{sweller1985use}~\cite{koedinger2012knowledge} into the curriculum as students learn more robustly from tasks which are interleaved with problem solving practice ~\cite{koedinger2012knowledge}.

In summary, we designed the \ac{HRE} course based on general cognitive psychological research on skill acquisition, which is supported by certain types of instructions and assignments as described. Since motivation is a central factor, we also took intentional steps to bolster student motivation in our course design as described in the following. 

\subsubsection{Supporting Students' Motivation in \ac{HRE} Course}
Since motivation is a key element in learning (Section~\ref{FIE:background:knowacqu}), we employed the following design principles in our course to enhance students’ motivation. In cases where higher levels of motivation are associated with greater cognitive engagement and learning~\cite{pintrich2003motivational}~\cite{litzinger2011engineering}, tasks and materials must cater to both personal and situational interest. The \ac{HRE}  tasks we present in the practical phase are stimulating and engaging exercises which are both novel and touch upon a variety of real-world challenges encountered in \ac{HRE} practice (e.g., finding control logic, retrieving a cryptographic key, etc.). The integration of the \ac{HRE} software \HAL into the course helps students learn \ac{HRE}  processes through the use of realistic graphical representations~\cite{TALE:2018:Wiesen} which should in turn lead to growing interest and involvement. By providing authentic \ac{HRE} tasks and making connections to students’ intended profession, the course design supports an increase in the perceived value of the learning experience which again leads to enhanced motivation~\cite{ambrose2010learning}. Students who believe they are able to solve a task are more highly motivated in terms of effort and persistence~\cite{pintrich2003motivational}. Thus, it is important to include tasks which are on an appropriate level of difficulty and allow students to use their prior knowledge and skills. We consequently designed the tasks within the project phase to fall within the range of competence achieved by the conclusion of the lecture phase.
 
In summary, \ac{HRE} is important in the field of cyber security and specialists with \ac{HRE} skills are in great demand. In addition to the lack of \ac{HRE} experts, there is also a lack of educational university-level \ac{HRE} programs. Thus, we developed a \ac{HRE} course by referring to prior cognitive psychological research, since research on the acquisition of \ac{HRE} skills cannot be found. In order to evaluate the effects of our course design on \ac{HRE} skill acquisition, we run two studies with participants on different levels of prior knowlegde. In the following, we present our research methods and materials. 

\section{Methods}
\label{fie:section:materials}

\subsection{Research Questions}
To observe participants’ skill acquisition as well as correlations with motivational and cognitive factors, we formulated the following research questions:
\begin{enumerate}
	\item Does students’ performance in solving \ac{HRE} tasks improve with increasing experience?
	\begin{enumerate}
	\item Does increasing experience result in students needing less time to solve \ac{HRE} tasks?
	\item Do students exhibit a higher probability of solving \ac{HRE} tasks as their experience grows?
	\end{enumerate}
	\item Are there differences between students with different levels of expertise (undergraduate and graduates) regarding the hypothesized improvements?
	\item Do the hypothesized improvements relate to particular aspects of intelligence and related cognitive abilities (e.g., processing speed), or prior experiences in relevant topics?
\end{enumerate}

\subsection{Participants}
The first study was conducted at a North American university with 20 students (mean age $M=23.5$, $SD=2.3$; 9 undergraduates) who were enrolled in programs in electrical engineering or computer science. The ethics board approved the study. The second study was conducted at a German university with 18 participants (mean age $M=23.1$, $SD=1.8$; 9 undergraduates) who studied cyber security or electrical engineering. The institutions were chosen based on their strong programs in cyber security, and computer engineering. Five participants were excluded because they did not complete all the tasks and the amount of data was not sufficient for analyses. Both studies were conducted in winter term of 18/19. In both studies, participants provided written informed consent and received monetary compensation for spending time on answering study related surveys and tests. We ensured privacy by randomly assigning pseudonyms to the participants of both studies. These pseudonyms were consistently used throughout all materials and procedures regarding the two studies. 

\subsection{Materials}
\label{fie:subsection:materials}
\subsubsection{Educational Environment}
The \ac{HRE} framework \HAL \cite{tdsc:2018:fyrbiak, wallat2019highway} served as the underlying educational environment for the projects of the practical phase. \HAL assists users in the reverse engineering of complex gate-level netlists and its extensibility allows for the development of custom plugins. In particular, \HAL employs an interactive \ac{GUI} to provide both textual and graph-based representation of the netlist under inspection. While the graph-based representation allows for detailed manual inspection and highlighting of the netlist and its components, an integrated Python shell provides an efficient approach to further interact with and process the netlist via aforementioned plugins.

\subsubsection{\ac{HRE} Projects}
The practical phase consisted of four projects, of which each contained the following subtasks: (1)~the reading of relevant scientific papers, (2)~pen \& paper exercises, and (3)~practical reverse engineering tasks. 

In the following, we describe the projects with special emphasis on the practical tasks which had to be solved with \HAL. All practical \ac{HRE} tasks were based on flat \ac{FPGA} netlists without any high-level information such as variable and signal names, comments, hierarchies, or module boundaries. The netlists were available in \acs{VHDL} and synthetized for the Xilinx Spartan-6 architecture\cite{XILINX:2010:spartan6}. They were composed of global input and output buffers, \acp{LUT} and Multiplexers for combinational logic, \acp{FF} for sequential logic -- hereafter all of them simply referred to as \textit{gates} -- and their interconnections.

\subsubsection*{Project~1 -- Introduction to Gate-level Netlist Reverse Engineering}
This project introduced the \HAL environment and its basic features to the students. In the practical task, students had to analyze the data path of an unknown substitution-permutation-network called \textit{ToyCipher}: they had to determine the block and key sizes of the cipher, to decide if the implementation was round-based or unrolled, and to identify the SBoxes. Due to the relatively low complexity of the design (131 gates) and the straightforwardness of the tasks, this assignment could be solved mostly through manual inspection of the netlist in \HAL.

\subsubsection*{Project~2 -- Control Logic Reverse Engineering}
In this project, students were directed to reverse engineer the control logic from a slightly modified variant of the \textit{ToyCipher} from project~1. Therefore, students identified the logic gates of which the \ac{FSM} implementing the control logic was composed via graph-based analysis and manual inspection of the candidates. While the basic functionality as well as the complexity (138 gates) of the underlying netlist was similar to the previous one, this assignment focused on the understanding and implementation of the methods used for semi-automated \ac{FSM} extraction.

\subsubsection*{Project~3 -- Reverse Engineering of Obfuscated Control Logic}
The underlying 128-gate netlist for this project was a second variant of the \textit{ToyCipher} utilizing the control flow obfuscation method Harpoon \cite{tcad:2009:chakraborty}. Obfuscation in this context is a transformation which obstructs high-level information without changing functionality~\cite{wiesen2019towards}. The goal of obfuscation is to impede the reverse engineering processes. Students had to extract the gates implementing the control logic and analyze the obfuscation method by differentiating the obfuscated and the original parts. In the second step, they disabled the obfuscation through initial state patching and verified their result through dynamic analysis of the netlist.
While the basic functionality as well as the complexity of the underlying netlist was similar to the previous one, this assignment focused on building understanding of obfuscated control logic as well as practicing dynamic analysis of netlists.

\subsubsection*{Project~4 -- \ac{AES} Key Extraction}
In the last project, students had to extract a hard-coded key from a netlist implementing a real-world \ac{AES} design. \ac{AES} is the most widely used encryption algorithm.
The first task was to derive high-level information such as the functionality (encryption or decryption), the presence of the key schedule, the key length, and the hardware architecture (round-based or unrolled) from this substantially more complex netlist (2176 gates). Secondly, they had to write a script to identify the Sbox logic, since the Sboxes served as a potential anchor for attacks on the hard-coded key. Finally, the hard-coded key had to be extracted through manipulation and dynamic analysis of the underlying circuit.
For this project, already learned \ac{HRE} techniques such as the derivation of high-level information, identification of functional blocks through scripting, and dynamic netlist analysis had to be applied in a significantly larger environment than before.

\subsection{Measures \& Instruments}
\subsubsection{Solution Time and Solution Probability}
In the studies, we focused on observing changes to and influences on two dependent variables which are traditional measures in cognitive psychology: time on task (solution time), and accuracy in the task (solution probability). In the following, we present how we measured them. 
\HAL automatically tracked participants' behavior through the creation of log files with time stamps for every interaction within \HAL (please note that no personal information was recorded). After providing informed consent, participants uploaded their pseudonymized log files. We calculated the solution time per project per student based on these log files. 
To calculate the solution time accurately, we set an inactivity threshold of $t = 1$ hour and subtract periods longer than $t$ from the total duration between start and finish of the projects. Analyses for solution time were conducted with data of 20 participants, since the data of the remaining participants was not continously available.
Every participant received a grading for the four \ac{HRE} projects, because the projects were embedded in an academic course. The resulting scores were the basis from which we calculated the per-project solution probabilities. The scores from every project were standardized as percentage to enable comparison. The analyses for solution probability were computed with the whole sample of 38 participants.

\subsubsection{Control Variables}
A self-developed questionnaire on socio-demographics asked participants to provide information about their age, major, and target degree. Additionally, students were requested to describe their prior experiences in relevant topics (e.g., Boolean algebra, \acp{FSM}, symmetric cryptography, Python programming, etc.) on a 5-point Likert-Scale, ranging from 1 (very low) to 5 (very high). Item scores were summed for analyses.

\subsubsection{Further Variables of Interest}
The \ac{WAIS-IV}~\cite{wechsler2008wechsler} was used to assess the students’ cognitive abilities. It consisted of ten tests to measure four sub scores: \ac{VC}, \ac{PR}, \ac{WM}, and \ac{PS}. \ac{VC} quantified abstract verbal reasoning and verbal expression abilities. It was assessed by the three tests: Similarities (participants were asked to describe how two words are similar), vocabulary (participants defined words), and information (participants answered questions about general knowledge). It should be noted, that students who were not native speakers of German did not complete tests on \ac{VC}. \ac{PR} measured the ability to accurately interpret and work with visual information. It consisted of three tests: Block design (participants rearranged 3-dimensional blocks to match patterns), matrix reasoning (participants completed 2-dimensional series of figures), and visual puzzles (participants chose three figures from which to build a 2-dimensional geometric shape). \ac{WM} reflected the ability to memorize information and to perform mental operations using that information. It consisted of two tests: Digit span (participants recalled a series of numbers in a given order), and arithmetic (participants solved arithmetical problems). \ac{PS} quantified the participants’ ability to process visual information quickly and efficiently. It consisted of two tests: Symbol search (participants were asked to search symbols rapidly and accurately), and coding (participants needed to transcribe a unique geometric symbol with its corresponding Arabic number rapidly and accurately). The analyses of the \ac{WAIS-IV} provided a \ac{FSIQ} based on the combined sub scores of \ac{VC}, \ac{PR}, \ac{WM}, and \ac{PS}.

To investigate the students’ level of motivation, we employed the \ac{QCM}~\cite{rheinberg2001qcm} during each of the four projects. The \ac{QCM} consisted of 18 items which measured the following four motivational factors on a five-point Likert scale from 1 (strongly disagree) to 5 (strongly agree): expected challenge of a task (“This task is a real challenge for me”), probability of success (“I think I am up to the difficulty of this task”), participants’ interest (“I would work on this task even in my free time”), and anxiety of failure (“I’m afraid I will make a fool out of myself”). Students answered the \ac{QCM} via the online survey provider Soscisurvey. After inverting items that were pooled differently, we computed means of the four sub factors.

As is commonly practiced in current research on Cognitive Load~\cite{schmeck2015measuring}, we integrated the Perceived Task Difficulty Scale~\cite{1972:bratfisch}, and the Mental Effort Scale~\cite{1992:paas}. The Cognitive Load Scales were used to determine if participants recognized the increasing complexity of the \ac{HRE} projects. Participants were asked to rate their Perceived Task Difficulty on a 7-point Likert Scale, ranging from 1 (very very easy) to 7 (very very difficult). Additionally, students rated their invested amount of mental effort on a 7-point Likert Scale, ranging from 1 (very very low) to 7 (very very high) via the online survey provider Soscisurvey. We computed the means of each scale.

\subsection{Study Procedure}
We conducted the quasi-experimental studies with a within-subject design during the winter term 2018/2019 at one German and one North American university. The studies were integrated into the practical phase of the \ac{HRE} course (Section~\ref{FIE:background:reforcd}). After students signed informed written consent documents and received a randomly-assigned pseudonym, the studies started with online questionnaires on socio-demographics, and prior experiences in relevant topics. Over the course of the semester, the \ac{WAIS-IV} was administered once in a 90-120 minute face-to-face session with each student. Overall, we used a similar procedure to collect data at four different points in time (four HRE projects) as described in the following. After reading the assignment of the current \ac{HRE} project, participants were asked to rate their level of current motivation (\ac{QCM}) regarding the imminent \ac{HRE} task. After finishing the task, students uploaded their log files of the current \ac{HRE} project to the SFTP server and subsequently answered the two Cognitive Load Scales on Mental Effort and Perceived Task Difficulty.

\section{Results}
\label{fie:section:results}
\begin{figure}[htb]
	\subfloat[]{\includegraphics[clip, width=.9\columnwidth]{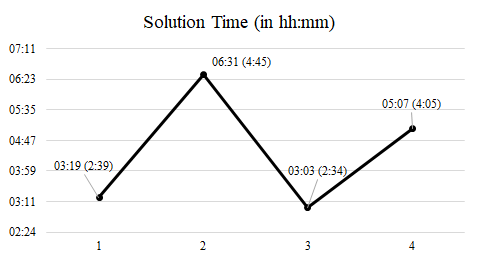}
		\label{fig:soltim}}
	
	\subfloat[]{\includegraphics[clip, width=.9\columnwidth]{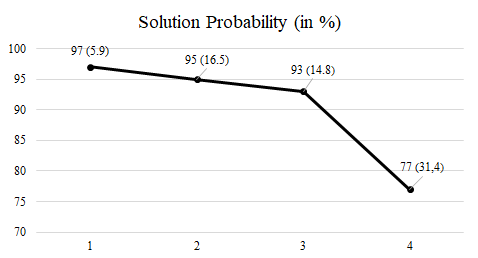}
		\label{fig:solprob}}
	\caption{Means (\textit{M}) and Standard Deviations (\textit{SD}) of solution time~(a) and solution probability~(b) in the format \textit{M (SD)}.}
	\label{fig:soltimprob}
\end{figure}

To answer research questions 1a and 1b, we conducted a repeated-measures ANOVA of solution times and probabilities which is a common method for comparing changes over time in psychological research (Fig.~\ref{fig:soltimprob}). Since, we did not find any group differences between students from both universities, we were able to merge the two samples in our analyses. The results showed significant differences across the four times of measure, with $F (3, 17) = 5.66$, \textit{p} = .03, $\eta^2$ = .50 (Fig.~\ref{fig:soltim}). The post-hoc analysis revealed that the solution time differed significantly between all projects, except for solution time between projects 1 and 3, and projects 2 and 4. 

The repeated-measures ANOVA for comparing the mean of solution probabilities across the four \ac{HRE} projects (Fig.~\ref{fig:solprob}) revealed that students’ solution probability decreased significantly in the most complex \ac{HRE} project~4, \textit{F} (3,35) =7.09, \textit{p}=.00, $\eta^2$ = .38. It should be noted, however, that the mean solution probability (\textit{M}=77.2, \textit{SD}=31.4) was still at a satisfactory level. We found no significant correlation between solution time and solution probability across the four projects.

To test, if students perceived the increasing complexity of the single projects, we conducted a repeated-measures ANOVA for the two Cognitive Load scales. The results showed that students reported a significant higher mental effort in project~4 (\textit{M}=4.26, \textit{SD}=1.59) compared to project~1 (\textit{M}=3.45, \textit{SD}=1.43) with, \textit{F} (3, 35) = 3.56, \textit{p} = .024, $\eta^2$ = .23. The repeated measures ANOVA for the Perceived Task Difficulty scale showed significant results between the means of project~2 (\textit{M}=3.61, \textit{SD}=1.29) compared to project~3 (\textit{M}=5.39, \textit{SD}=1.26), and compared to project~4 (\textit{M}=5.24, \textit{SD}=1.71), with \textit{F}(3, 35) = 18.77, \textit{p} = .000, $\eta^2$   = .62. These results showed that students perceived the growing complexity of the projects.

To test if our course design (Section~\ref{FIE:background:reforcd}) supported continuous motivation across the practical phase, we computed the repeated-measure ANOVA of the \acf{QCM}.
The analysis revealed no significant differences between the students’ levels of motivation across the four times of measure. The descriptive analyses showed above-average levels of motivation over all projects, with Interest (\textit{M}=3.9, \textit{SD}=0.38), Challenge (\textit{M}=3.9, \textit{SD}=0.46), Anxiety (\textit{M}=1.59, \textit{SD}=0.34), and Probability of Success (\textit{M}=3.65, \textit{SD}=0.24).

To answer research question 2, we computed the following analyses. To assess differences in the effect of the level of expertise on solution probability we calculated a repeated-measures ANOVA. The analysis revealed no significant effect of expertise on solution probability with (\textit{F} (3, 34) = 0.17, \textit{p} = .92,$\eta^2$ = .02), nor on solution time with (\textit{F} (3, 17) = .94, \textit{p} = .44, $\eta^2$ = .15).

We conducted bivariate correlations between \acf{FSIQ}, IQ sub factors and prior experiences in the light of research question~3. Table~\ref{tab:correlations} shows the results, which indicate significant correlations between the sub factors \acf{WM}, and \acf{PS}, and prior experiences in \acp{FSM}, symmetric cryptography, and students’ performances.
\begin{table}[htb!]
	\centering
	\caption[asdf]{Pearson correlation with correlation coefficients and significances.}
	\label{tab:correlations}
	\begin{tabular}{lllllllll}
		\hline
		& \multicolumn{4}{c}{Solution Probability} & \multicolumn{4}{c}{Solution Time} \\
		\hline
		& P1 & P2 & P3 & P4 & P1 & P2 & P3 & P4\\
		\hline
		\multicolumn{9}{c}{\textbf{IQ and sub factors}}\\
		\ac{FSIQ}$^\dagger$&.15 &.17&.12&.18&.19&.40&.32&.38 \\
		\ac{WM}$^\dagger$&.47&.10&.56&.72$^*$&.01&.19&.43$^*$&.07 \\
		\ac{PS}$^\dagger$&.10&.15&.16&.69$^*$&.03&.02&.59$^*$&.17 \\
		\ac{PR}$^\dagger$&.03&.14&.20&.31&.18&.46&.56&.88 \\
		\hline
		\multicolumn{9}{c}{\textbf{Prior Experiences}}\\
		\ac{HRE}&.10&.18&.20&.19&.09&.19&.25&.01\\
		Bool. Algeb.&.01&.06&.08&.13&.04&.21&.10&.06 \\
		\acp{FPGA}&.25&.07&.21&.01&.25&.29&.01&.03 \\
		\acp{FSM}&.01&.06&.14&.01&.18&.19&.42$^*$&.15\\
		Sym. Crypto&.35&.17&.16&.08&.41$^*$&.13&.14&.53$^*$\\
		\hline
		\multicolumn{9}{l}{$^*$Correlation is significant at the $0.05$ level.}\\
		\multicolumn{9}{l}{$^\dagger$\acf{FSIQ}, \acf{WM}, \acf{PS},}\\
		\multicolumn{9}{l}{\acf{PR}.}
		
	\end{tabular}
\end{table}

As no factor correlated significantly with time and probabilities across all four projects, the computation of ANCOVAS was not feasible. To explore potential differences between students with different levels of expertise, prior experiences, and cognitive abilities, we calculated repeated-measures ANOVAS. Each independent variable was transformed by a median split to conduct the calculations. In the following, we only report the calculations with independent variables which showed significant correlations (Table 1).

The repeated-measures ANOVA revealed significant differences between students with higher and lower \acf{WM} scores on solution time (\textit{F} (3, 17) = 1.27, \textit{p} = .04, $\eta^2$  = .19). Students with higher \ac{WM} scores were significantly faster in solving project~2 (\textit{p} = .03), 3 (\textit{p} = .04), and 4 (\textit{p} = .05). Additionally, the results revealed significant differences between a higher and lower \ac{WM} related to solution probability, with (\textit{F} (3, 34) = 5.85, \textit{p} = .05, $\eta^2$  = .34). Students with higher \ac{WM} scores were significantly better at solving project~4.

To assess the effects of \acf{PS}, we calculated a repeated-measures ANOVAs for solution time and probability. The results revealed significant differences between higher and lower \ac{PS} scores regarding solution probability (\textit{F} (3, 34) = 9.43, \textit{p} = .041, $\eta^2$ = .45). Students with higher \ac{PS} scores had a significant higher solution probability in project~4 (\textit{p} = .04) than students with lower \ac{PS} scores. Repeated-measures ANOVA for \ac{PS} and solution time revealed a significant effect (\textit{F} (3, 17) = 1.3, \textit{p} = .04, $\eta^2$  = .19). Students with higher scores in \ac{PS} solved project~3 faster than students with lower \ac{PS} scores. 

Additionally, the calculation for prior experiences showed significant effects. The repeated-measures ANOVA revealed significant differences between students with high and low levels of experiences in symmetric cryptography regarding solution time (\textit{F} (3, 17) = 4.41, \textit{p} = .02, $\eta^2$ = .45).  Students with more experience in symmetric cryptography solved projects 1 and 4 significantly faster than students with less experiences in symmetric cryptography.

The repeated-measures ANOVA did not reveal significant effects of higher and lower experiences in \acp{FSM} on solution time (\textit{F} (3, 17) = 1.8, \textit{p} = .19, $\eta^2$ = .25), nor on solution probability (\textit{F} (3, 34) = 0.9, \textit{p} = .43, $\eta^2$ = .07)


\section{Discussion}
\label{fie:section:discussion}

In the light of research question 1a, our data showed that students needed significantly more time for solving project 2 than for solving project 1. This is not a surprise, since students were asked to create automated solutions for the first time which might have been challenging and time consuming in comparison to project 1. Interestingly, our data showed that students were able to complete project 3 more quickly than project 2, despite project 3 being the more complex task. We here assume that memory and fluency-building are involved in the development of declarative and procedural \ac{HRE} knowledge~\cite{koedinger2012knowledge}. Students enrolled in our course were able to use their knowledge gained in project 2, for solving project 3 faster, thus demonstrating that they had acquired \ac{HRE} skills which enabled them to become more fluent. Overall, we observed a significant increase of solution time between projects 1 and 4, and an observable but not significant increase between projects 3 and 4, which account to the growing complexity of the underlying gate-level netlists. 

Referring to research question 2, the analyses revealed no significant differences between students with different levels of expertise in regard to solution time and solution probability. It would have been expected, that graduate students had acquired more relevant prior knowledge in important topics which might have resulted in shorter time on task and higher solution probabilities. Nevertheless, our results showed no differences between students on different levels of expertise. We conclude, that graduate students did not acquire more relevant knowledge for solving \ac{HRE} tasks in previous university courses than undergraduates did. Our results show, that our course taught a sufficient amount of relevant knowledge during the lecture phase which equally enabled both undergraduate and graduate students to solve the \ac{HRE} projects.

In the light of research question 3 we conducted several computations to observe which cognitive factors or prior knowledge had a significant effect upon solution probability and solution time. \acf{WM} is one important cognitive factor in acquiring \ac{HRE} skills. Our data showed that higher \ac{WM} scores led to faster solutions in projects 2, 3, 4, and higher solution probability in  project~4. Referring to the key characteristics of \ac{WM} helps to illustrate how a good \ac{WM} is important to the successful completion of \ac{HRE} projects. \ac{WM} is defined as a system for storing and manipulating information in the context of complex tasks such as learning and problem solving~\cite{baddeley1974working}. It enables the retrieval of learned information from long term memory~\cite{dehn2011working}, and keeps both relevant old and novel facts in memory during the activity of problem solving while at the same time ignoring irrelevant information~\cite{hill2010can}. By taking this into account, we can explain why students with higher \ac{WM} scores solved the \ac{HRE} projects more quickly -- namely because they were better able to combine novel and stored information and ignore irrelevant information, thereby increasing the speed with which they reached solutions in the individual projects. A good \ac{WM} also supported students in reaching higher solution probabilities in project~4. Briefly revisiting the specific requirement of that project makes it rather obvious why that was the case: to successfully complete the project, students had to recall knowledge, work flows, and problem-solving strategies that they had learned during projects~1-3. For example, during project~4, students had to recall the derivation of high-level information and Sbox identification skills taught in project~1, as well as the dynamic netlist analysis competencies fostered during project~3. Students with a good \ac{WM} were supported in recalling relevant information that they learned earlier in the course, leading them to achieve higher solution probabilities and shorter solution times.  

\acf{PS} is another relevant cognitive factor in the context of acquiring \ac{HRE} skills as demonstrated by the observation that students with higher \ac{PS} scores had a significantly higher solution probability in project~4 and solved project~3 significantly faster than did students with lower \ac{PS} scores. \ac{PS} is defined as the ability to process visual information quickly and efficiently, and research on \ac{PS} has shown that it can be a good predictor of how quickly and accurately students can perform a task~\cite{lichtenberger2009essentials}. Higher \ac{PS} scores therefore contributed to solving project~4 more accurately and project~3 more quickly. 

In the context of research question~3, our analysis established the significant effects that prior experience with \acp{FSM} and symmetric cryptography had upon student outcomes, with students who had more experience in both areas performing better than students with less experience. Our data showed a significant correlation between prior experiences in \acp{FSM} and solution time in project 3. The twofold challenge of project 3 consisted of first detecting an \ac{FSM} and second breaking the \ac{FSM} obfuscation. A higher prior knowledge in \acp{FSM} might have been helpful to solve the first challenge more quickly, since students did not need to acquire knowledge on \acp{FSM} during their work on project 3. Additionally, our analyses showed that prior knowledge in symmetric cryptography enabled students to solve projects 1 and 4 more quickly. Prior knowledge on the inner workings of symmetric ciphers is advantageous regarding the contents of projects 1 and 4, e. g. data path analysis, the detection of Sboxes, or general attack strategies on such ciphers.

\subsection{Implications for Future \ac{HRE} Course Designs}

The integration of Spacing and Testing ~\cite{pashler2007organizing,koedinger2012knowledge} throughout the four \ac{HRE} projects supported students’ development of memory and fluency, and the inclusion of Worked Examples ~\cite{sweller1985use,koedinger2012knowledge} enabled students to acquire \ac{HRE} skills from tasks with problem solving practice. These two instructional principles should be part of future \ac{HRE} courses. We also assume that the integration of stimulating and realistic exercises of an appropriate difficulty level as well as the integration of \HAL led to the above-average levels of student motivation we observed during the course and should be included in future \ac{HRE} courses. 
We found significant differences between students on different levels of prior knowledge. To compensate these differences, future \ac{HRE} courses should include special programs for students with lower levels of prior knowledge in relevant topics, e.g., basic knowledge lectures and exercises regarding symmetric cryptography or tutorials for practicing Python programming, since \ac{HRE} is automatable in \HAL via Python. 
Our results revealed significant differences between students with higher and lower scores in \acf{WM}. Due to time factors, a course on \ac{HRE} cannot not include training to improve \ac{WM} performance. Nevertheless, we could structure the \ac{HRE} course in a way, which supports students with lower scores in \ac{WM}. By referring to the fact that the capacity of \ac{WM} is limited, we can design our assignments and projects by supportive knowledge, students already learned in earlier stages of the course. This could be a possible way to support students with lower scores in \ac{WM} to focus on the current task instead of struggling in recalling information from prior lectures and projects. 

\subsection{Limitations and Implications for Future Research}
This presented work has limitations that should be investigated. In the future, studies on human processes in \ac{HRE} with a larger sample size would be preferable to produce results with a higher impact and significance. Since this study is a first investigation to fill the research gap of skill acquisition in \ac{HRE}, the generalizability to other areas is limited. 
Our research prompted us to make several observations about potential future research on skill acquisition in the field of \ac{HRE}. It would be preferable to further analyze problems, errors, or difficulties of human reverse engineers (e.g., process modelling of applied problem solving strategies). Doing so would allow us to offer specific support via instructions, exercises, or training. If \acf{WM} and \acf{PS} are relevant factors in predicting expertise development, a differential examination of the central executive~\cite{baddeley1996fractionation} might shed more light on the role of cognitive factors in acquiring \ac{HRE} skills. Since our data led us to assume that declarative knowledge had been transformed into procedural knowledge, a closer examination of the types of knowledge presented in the two phases of the course might prove interesting. Finally, the integration of a second complex task into a future study would help reveal more individual differences between students as well as help ascertain whether differences in solution time and solution probability are stable over a longer time horizon consisting of multiple complex projects.

\section{Conclusion}
\label{fie:section:conclusion}
\ac{HRE} is important in the field of cyber security and therefore \ac{HRE} skills are in high demand across industries and around the world. Despite the clear and unmet worldwide demand for \ac{HRE} experts, there is a surprising lack of educational \ac{HRE} courses, and research on the acquisition of \ac{HRE} skills is lacking thus far. Against this background, we developed a \ac{HRE} course based on psychological cognitive research and augmented through the integration of specific instructional methods which are known to support the development of declarative and procedural knowledge. Through the conduction of two quasi-experimental studies, we demonstrated that our course supported the acquisition of \ac{HRE} skills. Our students demonstrated increased fluency in \ac{HRE} skills by solving increasingly complex tasks in successively shorter time intervals. Statistical analyses established that differences in individual student outcomes were a function of differences in \acf{WM}, \acf{PS}, and prior knowledge in relevant topics. Finally, we derived ideas for future course designs and research aimed at achieving a deeper understanding of the underlying psychological factors behind the acquisition of \ac{HRE} skills, such as observing the central executive of the \ac{WM} or by observing the impact that the integration of further complex tasks has upon solution time and solution probability.

\bibliographystyle{IEEEtran}
{
	\footnotesize \bibliography{IEEEabrv,bibliography}
}

\end{document}